\documentstyle[aps,epsfig,twocolumn]{revtex}

\begin{document}
\title{Arbitrary-speed quantum gates within large ion crystals through minimum control of laser beams}
\author{Shi-Liang Zhu, C. Monroe, and L.-M. Duan}
\address{FOCUS center and MCTP, Department of Physics,
University of Michigan, Ann Arbor, MI 48109 } \maketitle

\begin{abstract}
We propose a scheme to implement arbitrary-speed quantum entangling gates on
two trapped ions immersed in a large linear crystal of ions, with minimal control of laser beams. For gate speeds slower than the oscillation frequencies in the trap, a single appropriately-detuned laser pulse is sufficient for
high-fidelity gates. For gate speeds comparable to or faster than the local
ion oscillation frequency, we discover a five-pulse protocol that exploits
only the local phonon modes. This points to a method for efficiently scaling
the ion trap quantum computer without shuttling ions.
\end{abstract}

Significant advances have been made towards trapped ion quantum
computation in the last decade
\cite{1,2,3,4,5,5',6,7,8,9,10,11,12,13,14,15,16,Staanum,17,18}.
Many ingredients of quantum computing have been demonstrated
experimentally with this system
\cite{6,7,8,9,10,11,12,13,14,15,16}; and different versions of
quantum gate schemes have been proposed, each offering particular
advantages \cite{2,3,4,5,5',Staanum,17,18}. In conventional
approaches to trapped ion quantum gates, the interaction between
the ions is mediated by a particular phonon mode (PM) in the ion
crystal through the sideband addressing with laser beams. In these
types of gates, the control of the laser beams is relatively
simple, requiring only a continuous-wave beam with an appropriate
detuning; but to resolve individual motional sidebands, the gate
speed must be much smaller than the ion trap oscillation
frequencies. More recently, fast quantum gates have also been
proposed, which can operate with a speed comparable with or
greater than the trap frequencies \cite{17,18}. These types of
gates involve simultaneous excitation of all PMs
\cite{16,Staanum,17,18} and require more complicated control of
either the pulse shape \cite{17} and/or the timing of a fast pulse
sequence\cite{17,18}.

In this paper, we develop a gate scheme that combines the
desirable features of the above two types of gates. A conditional
phase gate with arbitrary speed is constructed in a large ion
array by optimization of few relevant experimental parameters. As
a result, first we show that with simple control of the detuning
of a continuous-wave laser beam, one can achieve a high fidelity
gate with the gate speed approaching the ion trap frequency. This
result is a bit surprising as many PMs are excited during the
gate. However, with control of just one experimental parameter
(the detuning), each of the modes becomes nearly disentangled with
the ion internal states at the end of the gate. Secondly, we show
that as the gate speed becomes larger than the local ion
oscillation frequency (specified below and see also Ref.
\cite{18}), only ``local" PMs will be primarily excited during the
gate. This yields a scaling method for trapped ion quantum
computation: a significant scaling obstacle to trapped ion quantum
computation is that due to the long range Coulomb interaction, any
collective gate on two ions is necessarily influenced by all the
other ions in the architecture, which makes the gate control
increasingly difficult with growth of the qubit number.
Conventionally, one needs to use the ion shuttling in a
complicated trap architecture to avoid this undesirable influence
\cite{5,6,9,14}. However, if the gate speed becomes comparable
with the local ion oscillation frequency, we have an alternative
scaling method without the requirement of ion shuttling: one can
perform the gate by exciting only the local PMs, which avoids the
complicated influence from the background ions. This result also
improves the fast gate scaling method proposed in Ref. \cite{18},
as here to excite only the local PMs, instead of using hundreds of
short pulses, we only need to apply five long pulses with
optimized amplitudes chopped from a continuous wave laser beam.

The system we have in mind is $N$ ions in a linear trap with a global trap
frequency $\omega$. To perform arbitrary-speed quantum gates, we need to
consider all the PMs \cite{17,18}. Without laser beams, the ion motional
Hamiltonian has the standard form $H_{0}=\sum_{k=1}^{N}\hbar \omega
_{k}(a_{k}^{\dagger }a_{k}+1/2)$ with $a_{k},a_{k}^{\dagger }$ as the
annihilation and creation operators of the $k$th PM. The eigen-frequency of
the PM $\omega _{k}\equiv \sqrt{\mu _{k}}\omega $ is determined by solving
the eigen-equations $\sum_{n}A_{nl}{\bf b}_{n}^{k}=\mu _{k}{\bf b}_{l}^{k}$,
where the matrix elements $A_{nl}=1+2\sum_{p=1,p\not=%
l}^{N}1/|u_{l}-u_{p}|^{3}$ for $n=l$, and $A_{nl}=-2/|u_{l}-u_{n}|^{3}$ for $%
n\not=l$. The parameter $u_{n}=x_{n}^{0}/\sqrt[3]{e^{2}/4\pi \epsilon
_{0}M\omega ^{2}}$ with $x_{n}^{0}$ representing the equilibrium position of
$n$th ion and $M$ denoting the mass \cite{19}. To perform quantum gates, we
need to apply some spin-dependent force on the ions, which can be induced,
for instance, through the ac-Stark shift from two propagating laser beams
with a relative angle and detuning \cite{11}. As it is the case in
experiments \cite{11,15}, we assume that when the ions are in their
equilibrium positions, the ac-Stark shifts for the ion qubit states $\left|
0\right\rangle $ and $\left| 1\right\rangle $ are equivalent.
Then, under the Lamb-Dicke condition and in the interaction picture with
respect to $H_{0}$, the Hamiltonian for the spin-dependent force is given by
\begin{equation}
H=-\sum_{n,k=1}^{N}F_{n}(t)g_{n}^{k}(a_{k}^{\dagger }e^{i\omega
_{k}t}+a_{k}e^{-i\omega _{k}t})\sigma _{n}^{z},  \label{H_int}
\end{equation}
where $\sigma _{n}^{z}\equiv \left| 1\right\rangle \left\langle 1\right|
-\left| 0\right\rangle \left\langle 0\right| $ is the Pauli operator, $%
F_{n}(t)$ is the force on the $n$th ion, and $g_{n}^{k}=\sqrt{\hbar
/2M\omega _{k}}{\bf b}_{n}^{k}$ is the coupling constant between the $n$th
ion and the $k$th PM. Using the Magnus' formula, the evolution operator
corresponding to the Hamiltonian (1) is found as \cite{20}
\begin{equation}
U(\tau )=\exp [i\sum_{n}\phi _{n}(\tau )\sigma _{n}^{z}+i\sum_{l,n}\phi
_{ln}(\tau )\sigma _{l}^{z}\sigma _{n}^{z}],  \label{U2}
\end{equation}
where $\phi _{n}(\tau )=\sum_{k}[\alpha _{n}^{k}(\tau )a_{k}^{\dagger
}-\alpha _{n}^{k\ast }(\tau )a_{k}]$ with $\alpha _{n}^{k}(\tau )=\frac{i}{%
\hbar }\int_{0}^{\tau }F_{n}(t)g_{n}^{k}e^{i\omega _{k}t}dt,$ and $\phi
_{ln}(\tau )=\frac{2}{\hbar ^{2}}\int_{0}^{\tau
}\int_{0}^{t_{2}}\sum_{k}F_{l}(t_{2})g_{l}^{k}g_{n}^{k}F_{n}(t_{1})\sin
\omega _{k}(t_{2}-t_{1})dt_{1}dt_{2}.$

A conditional phase flip (CPF) gate on arbitrary two ions $i$ and
$j$ can be accomplished with identical spin-dependent forces on
only these two ions with $F_{i}(t)=F_{j}(t)=F(t)$. In this case,
the evolution operator $ U(\tau )$ in Eq. (2) exactly corresponds
to a CPF gate $U_{ij}=\exp (i\pi\sigma _{i}^{z}\sigma _{j}^{z}/4)$
if $\phi _{ij}(\tau )=\pi /4$ and $\alpha_{i(j)}^{k}(\tau )=0$ for
all the modes $k$. In principle, it is always possible to satisfy
this set of constraints by designing a sufficiently complicated
pulse shape for the forces. However, this kind of solution
typically requires exquisite control of many parameters that
determine the exact shape of $F(t)$, which may be difficult
experimentally. In the following, we show that in typical cases it
is only necessary to approximately satisfy these constraints,
allowing a much simpler class of laser pulse shapes to be used.

To design gate, we optimize the gate fidelity subject to a certain
class of laser pulses, with simple control parameters. With an
initial state $|\Psi _{0}\rangle $, the final state would be given
by $|\Psi _{f}\rangle =U_{ij}|\Psi _{0}\rangle $ after a perfect
CPF\ gate. However, with imperfect control, some of the PMs will
not evolve along a closed loop in the phase space corresponding to
$\alpha _{i,j}^{k}(\tau )\neq 0$. In that case, the final internal
state of the ions is mixed and described by the density operator
$\rho _{r}=Tr_{m}[U(\tau )|\Psi _{0}\rangle \langle \psi
_{0}|U^{\dagger }(\tau )],$ where the trace is over the motional
state of all the ions. The overlap between the ideal state $|\Psi
_{f}\rangle $ and the actual density operator $\rho _{r}$ defines
the fidelity $F_{g}=\langle \Psi _{f}|\rho _{r}|\Psi _{f}\rangle
$.
Without loss of generality, we choose here a typical initial state with $%
|\Psi _{0}\rangle =\left( \left| 0\right\rangle _{i}+\left| 1\right\rangle
_{i}\right) \otimes \left( \left| 0\right\rangle _{j}+\left| 1\right\rangle
_{j}\right) /2$ for calculation of the gate fidelity $F_{g}$. We assume that
the PMs are initially in thermal states with an effective temperature $T$.
Then, with the evolution operator $U(\tau )$ given in Eq. (2), the gate
fidelity $F_{g}$ is found to be
\begin{equation}
F_{g}=\frac{1}{8}[2(\Gamma _{i}+\Gamma _{j})+\Gamma _{+}+\Gamma _{-}],
\label{Infid}
\end{equation}
where $\Gamma _{i(j)}=\exp [-\sum_{k}|\alpha _{i(j)}^{k}(\tau )|^{2}\bar{%
\beta}_{k}/2]$, and $\Gamma _{\pm }=\exp [-\sum_{k}|\alpha _{i}^{k}(\tau
)\pm \alpha _{j}^{k}(\tau )|^{2}\bar{\beta}_{k}/2]$. The parameter $\bar{%
\beta}_{k}$ is given by $\bar{\beta}_{k}=\coth (\hbar \omega
_{k}/k_{B}T)=\coth [\frac{\sqrt{\mu _{k}}}{2}ln(1+1/\bar{n}_{c})]$, with $%
k_{B}$ denoting the Boltzman constant and $\bar{n}_{c}=(e^{\hbar \omega
/k_{B}T}-1)^{-1}$ representing the mean phonon number of the center-of-mass
mode.

To maximize the gate fidelity, we choose our control parameters to be simply the
detuning and the amplitude of the laser beams that introduce the spin-dependent force. With the ac Stark shift from the Raman laser beams \cite
{11,15}, the force function $F(t)$ has the form of $\Omega \sin (\mu t)$,
where $\mu $ is determined by the detuning between the Raman laser beams and
$\Omega $ is the two-photon Rabi frequency. To introduce more control
parameters, we can chop the continuous-wave laser beam into $m$ equal-time
segments with the Rabi frequency for the $p$th ($p=1,2,\cdots ,m$) segment
given by a controllable value $\Omega _{p}$. The force $F(t)$ then takes the
form $F(t)=\Omega _{p}\sin (\mu t)$ for the interval $(p-1)\tau /m\leq
t<p\tau /m$. This kind of amplitude control for the Raman beams can be done, for example, with simple acoustic- or electro-optical modulators.

With a sufficient number of control parameters $\Omega _{p}$, it is always
possible to make $F_{in}=0$. In this case, the conditions $\alpha
_{i,j}^{k}(\tau )=0$ require $\sum_{p=1}^{m}\Omega _{p}\int_{(p-1)\tau
/m}^{p\tau /m}\sin (\mu t)e^{i\omega _{k}t}dt=0$ for any $k$ mode, which are
a set of linear constraints for the ratios $f_{p}\equiv \Omega _{p}/\Omega
_{1}$. For the case of $N$ PMs, it is possible to satisfy these $N$ complex
constraints with $2N$ real parameters $f_{i}$ ($i=2,3,\cdots ,2N+1$), so the
required number of segments is $m=2N+1$. In the following, we will show that
we can actually use a much smaller number of segments (control parameters) to
reduce the gate infidelity to almost zero.

First we consider the case of the minimal control of the laser beams: a single
amplitude and detuning of the laser beam, or a single segment ($m=1$). This situation corresponds
exactly to current experimental configurations \cite{11,15}. Without
shape control of the laser beams, all the known gate schemes require the
gate speed to be much smaller than the ion trap frequency for sideband
addressing of a particular PM. Here, by taking into account of all the PMs,
we show that one can still get a high fidelity gate even if the gate speed
approaches the ion trap frequency, which is well beyond the limit set by the
sideband addressing.

In our calculation, we first consider the gate acting on the two central ions in a $20$-ion array. In Fig.1a, the gate fidelity calculated from Eq.(3) is shown
as a function of the laser detuning for various gate speeds. When the gate
time $\tau $ is significantly larger than $\tau _{0}$ with $\tau _{0}\equiv
2\pi /\omega $, the gate fidelity has local maximum at the detuning $\mu
=\omega _{k}+2\pi l/\tau $ with an integer $l$. This corresponds to the well
known condition in the phase (the Milburn-Sorensen-Molmer) gate \cite{3,4,11}%
. When $\tau $ approaches $\tau _{0}$, it is better to choose a detuning
with either $\mu <\omega $ or $\mu >\max \left\{ \omega _{k}\right\} $ to
have a higher gate fidelity. The optimal detunings shift a bit downwards in
the region $\mu <\omega $ and upwards in the region $\mu >\max \left\{
\omega _{k}\right\} $ compared with the values given by $\mu =\omega
_{k}+2\pi l/\tau $. A distinct result from this calculation is that the gate
fidelity can be still very high even if the gate speed goes well beyond the
sideband addressing limit. For instance, the optimal fidelity $F_{g}\simeq
99.97\%$ ($F_{g}\simeq 99\%$) for the gate time $\tau =2\tau _{0}$ ($\tau
=1.5\tau _{0}$), respectively, with the corresponding detuning $\mu $ pretty
close to $\omega -2\pi /\tau $. If we further increase the gate speed, the
fidelity quickly goes down. For instance, the optimal fidelity is only $80\%$
for $\tau =\tau _{0}$ and reduces to the minimum of $25\%$ (corresponding to
a completely mixed state after the gate) when $\tau \leq 0.05\tau _{0}$.

As the gate time $\tau $ approaches $\tau _{0}$, many PMs are
involved during the gate, and they ultimately get nearly
disentangled with the ion internal state. To see this, we checked
the contribution to the conditional phase $\phi_{ij}$ from all the
other (non center-of-mass) PMs for the case of the optimal
detuning very close to $\omega-2\pi /\tau $. The relative
contributions from the ``spectator" PMs is about $1.3\%$, $10\%$,
and $18.1\%$ for the gate time $\tau =50\tau _{0}$, $5\tau _{0},$
and $2\tau _{0}$, respectively. It is a bit surprising that, for
instance at $\tau =2\tau _{0}$, the spectator PMs contribute
$18.1\%$ of the conditional phase but induce an infidelity of only
$0.03\%$.

\begin{figure}[tbph]
\centering \epsfig{file=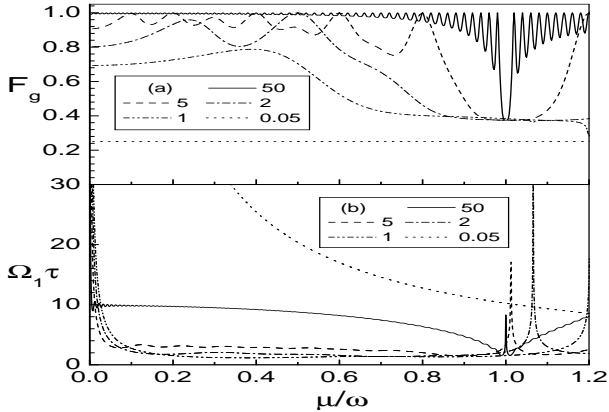,height=6cm,width=8cm} \caption{
For the two center ions in a $20$-ion array, the gate fidelity (a)
and the required Rabi frequency (b) shown as a function of the detuning $%
\protect\mu $ with $\protect\tau =50\protect\tau _{0},5\protect\tau _{0}, 2%
\protect\tau _{0},\protect\tau _{0},0.05\protect\tau _{0}$,
respectively. The other parameters: $\bar{n}_c=3$ and $m=1$.}
\label{Fig1}
\end{figure}

\begin{figure}[htbp]
\centering \epsfig{file=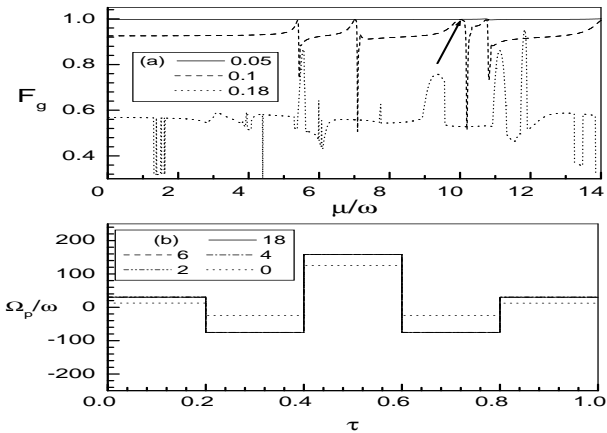,height=6cm,width=8cm} \caption{(a)
The gate fidelity as a function of the detuning $\protect\mu$
with $m=5$ and $\protect\tau=0.18\protect\tau_0,0.1\protect\tau_0,0.05%
\protect\tau_0$, respectively. (b)The optimal sequence of the force ($%
\Omega_p$) for the gate with $\protect\tau=0.1\protect\tau_0 $ and $\protect%
\mu=10\protect\omega$ (denoted by an arrow in (a)). The numbers
$n$ above the curve denote how many neighboring ions' motion is
taken into account for calculating the force sequence. With
$n=18$, all the PMs are included. The force sequences are
basically indistinguishable for $n=2,4,6,18$.} \label{Fig2}
\end{figure}

We have also calculated the required laser power (proportional to $\Omega
_{p}$ for the Raman configuration) for achieving the high-speed gates, and
the result is shown in Fig.1b. Note that the optimal detuning $\mu $
not only maximizes the gate fidelity, but it also requires the least amount of laser power. We can see from this figure that with increase of the gate speed, the required laser power grows slower than a linear increase in the region $\tau
\geq \tau _{0}$. In current experiments, typically $\tau \sim 100\tau _{0}$
\cite{11}, so with moderate increase of the laser power, one can expect a
significant increase of the gate speed even without any laser shape control.
Similar calculations are also done for gates on different pairs of ions in
the array. The results are qualitatively similar, although the gate fidelity
is somewhat lower for the pair of ions with a larger distance. For instance,
with $\tau =2\tau _{0}$ and an optimal $\mu $, the gate fidelity $F_{g}$ is
about $99\%$ for the 1st and 2nd ions (at the edge of the trap), and is $95\%
$ for the 1st and 20th ions (the worst case).

The above calculation shows that without chopping of the laser beams, the
gate fidelity quickly decrease in the region $\tau \leq 1.5\tau _{0}$. To
improve the gate fidelity, we need to introduce more control parameters by
dividing the laser beams into $m$ segments. One might expect that to attain
a certain gate fidelity, the number of segments $m$ (i.e., the number of
control parameters) should continuously increase with $1/\tau $, as more and
more normal PMs will be substantially excited during such a fast gate.
However, this is actually not the case as we will see here. The key point is
that as the gate speed becomes faster than the ion motional response time, only the local PMs (which are superpositions of many normal PMs) of the two ions involved in the gate will be substantially excited, greatly simplifying the control. To make this point more
precise, we characterize the ion response time by its local oscillation
frequency $\omega _{Li}$ \cite{18}. The $\omega _{Li}$ for the $i$th ion is
defined as the eigen-oscillation frequency of this ion if we fix all the
other ions in the trap at their equilibrium positions. If the gate speed becomes
faster than $\omega _{Li}$, we expect that the gate in a large ion array
could be reduced to an effective two ion problem, so with $m=2N+1=5$
segments of laser pulses, we should expect good gate fidelity. In the
following, we test this idea by calculating the gate fidelity with $5$ laser
pulses under various gate speeds.

\begin{figure}[tbph]
\centering \epsfig{file=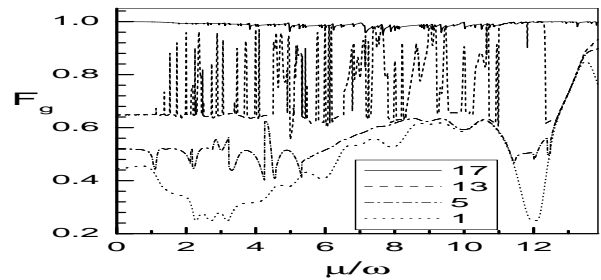,height=4cm,width=8cm} \caption{
The gate Fidelity as a function of the detuning $\protect\mu $
with $\protect\tau =0.5\protect\tau _{0}$ and $m =1,5,13,17$,
respectively. The other parameters are the same as in Fig. 1}
\label{Fig3}
\end{figure}

We still take a $20$-ion array, and for the center ions the local ion
oscillation frequency $\omega _{Li}\simeq 9.2\omega $. We calculate the gate
fidelity with $m=5$ and the optimized parameters $\Omega _{1},\Omega
_{2},\cdots ,\Omega _{5}$, and the result is shown in Fig.2a for $\tau
=0.18\tau _{0}$, $0.1\tau _{0}$, and $0.05\tau _{0}$, respectively. The
fidelity $F_{g}$ is above $99.99\%$ for all $\mu $ for $\tau \leq 0.05\tau
_{0}\sim 2\pi /\left( 2\omega _{Li}\right) $, clearly demonstrating the
above idea. Even for $\tau =0.1\tau _{0}\sim 2\pi /\omega _{Li}$,
we find that the gate fidelity is above $99\%$ at the optimal values of the
detuning with $\mu =5.4\omega $, $7.0\omega $, $10.0\omega ,$ or $10.7\omega
$. The corresponding force sequence $\Omega _{p}$ for $\mu =10.0\omega $
(corresponding to a fidelity $F_{g}=99.76\%$) is shown in Fig. 2(b). To see
that only the local PMs are substantially involved during the gate, we also
calculate the optimal $\Omega _{p}$ subject to the constraint that only a few
neighboring ions around the target ions are allowed to oscillate during the gate
(all the other ions are assumed fixed at their equilibrium positions).
Including the motion of $n$ ($n=0,2,4,6$) neighboring ions, the
corresponding optimal force sequences are shown in Fig. 2(b). The force
sequences become indistinguishable as soon as $n\geq 2$, which means that
the motion of the ions beyond the nearest neighbors has no influence on the
gate with a speed faster than $\omega _{Li}$. This result has important
implications for using the fast gates as a method to scale up ion trap
quantum computation \cite{18}. For a large scale computation with any ion
trap architecture, as soon as the gate speed becomes larger than the local
ion oscillation frequency, we need only consider the influence of neighboring ions on the target ions.
The other ions, near their equilibrium positions, only provide an effective static potential, and the design of the gate can always be well-approximated by considering only a few ions. So the control complexity of each gate does not increase with the number of ions in the computation, which provides an effective scaling method.

It turns out that it is most difficult to perform a gate with the
gate speed between the trap frequency $\omega $ and the local ion
oscillation frequency $\omega _{Li}$. In that region, one needs to
introduce more control parameters by dividing the laser beams into
more segments. But even in this worst case, it is still possible
to get a high-fidelity gate with the number of segments $m$ much
smaller than $2N+1$. For instance, with $20$ ions, the worst case
occurs with a gate time $\tau \sim 0.5\tau _{0}$, which requires
the largest number of control parameters. For this worst case, we
plot the gate fidelity in Fig.3 as a function of $\mu $ with
$m=1,5,13,17$, respectively. The fidelity $F_{g}$ has been above
$99\%$ at some optimal values of the detuning $\mu $ with $m=13$,
and a fidelity larger than $98.5\%$ can be reached at almost any
$\mu $ with $m\geq 17$. Note that this value is still significant
smaller than $2N+1=41$. We have also done calculation with
different number of ions in the array and for gates on different
pairs of ions. The results are qualitatively very similar to what
we have described. For instance,with $N=40$ ions, a gate fidelity
higher than $98.8\%$ can be reached for the two center ions with
the number of segments $m=1$ if the gate time $\tau \geq $
$1.7\tau _{0}$; and a fidelity large than $99.6\%$ is achievable
with $m=5$ segments of laser beams if the gate time $\tau \leq
2\pi /\omega _{Li}$ (in this case $\omega _{Li}=16.7\omega $ for
the center ions).

In summary, we have described a scheme to achieve arbitrary-speed quantum
gates on ions immersed in a large ion array, through minimum control of the amplitude of a continuous wave laser beam. With the same control complexity as the conventional gates, we have shown how to push the gate speed towards the ion trap frequency. We have also shown a version of fast gates with five laser pulses which can operator in any large ion crystal and thus provide an efficient
scaling method for ion trap quantum computation.

This work was supported by the ARDA under ARO contracts, the FOCUS center
and the MCTP, the NSF awards (0431476), and the A. P. Sloan Fellowship.
S.L.Z. was supported by the NCET.

\end{document}